\documentclass[jkps,fleqn,preprint,showpacs,showkeys]{revtex4}
\usepackage{graphicx}
\usepackage{amssymb}
\usepackage{amsmath}
\usepackage{bm}
\begin{document}
\setcounter{page}{0}
\title[]{A study on the performance of similarity indices and its relationship with link prediction: a two-state random network case}
\author{Min-Woo \surname{Ahn}}
\affiliation{Department of Physics, Pohang University of Science and Technology, Pohang, 37673, Republic of Korea}
\author{Woo-Sung \surname{Jung}}
\email{wsjung@postech.ac.kr}
\affiliation{Department of Physics, Pohang University of Science and Technology, Pohang, 37673, Republic of Korea}
\affiliation{Department of Industrial and Management Engineering, Pohang University of Science and Technology, Pohang, 37673, Republic of Korea}
\affiliation{Asia Pacific Center for Theoretical Physics, Pohang, 37673, Republic of Korea}

\date[]{}

\begin{abstract}
Similarity index measures the topological proximity of node pairs in a complex network.
Numerous similarity indices have been defined and investigated, but the dependency of structure on the performance of similarity indices has not been sufficiently investigated.
In this study, we investigated the relationship between the performance of similarity indices and structural properties of a network by employing a two-state random network.
A node in a two-state network has binary types that are initially given, and a connection probability is determined from the state of the node pair.
The performance of similarity indices affects the number of links and the ratio of intra-connections to inter-connections.
Similarity indices have different characteristics depending on their type. Local indices perform well in small-size networks and do not depend on whether the structure is intra-dominant or inter-dominant. In contrast, global indices perform better in large-size networks, and some such indices do not perform well in an inter-dominant structure.
We also found that link prediction performance and the performance of similarity are correlated in both model networks and empirical networks.
This relationship implies that link prediction performance can be used as an approximation for the performance of the similarity index when metadata for node types are unavailable.
This relationship may help to find the appropriate index for given networks.

\end{abstract}



\maketitle

\section{INTRODUCTION}
Complex network analysis has been an important framework for understanding the underlying structure of many systems.
Social interactions\cite{social1, social2, social3, social4}, economic relations\cite{econo1,econo2}, and biological and technical systems\cite{bio1,bio2,web1,web2} are composed of complex networks and are analyzed via complex network analysis.
Understanding their structure via empirical data has previously been investigated.
However, obtained networks are difficult to observe in their raw structure because of their huge size.
Thus, network parameters such as clustering coefficient and average path length are employed as representative parameters for the whole network structure\cite{WS, reviewnewman, reviewbarabasi}, or centrality measures such as betweenness centrality\cite{betweenness} and eigenvector centrality\cite{pagerank} are used as characteristics of individual nodes.
Similarity index is such a type of measure; it measures the structural proximity of a node pair.

The similarity index is employed to find similar nodes in empirical networks, such as node pairs that have a similar function in an empirical system (e.g., synonym extraction from a word network\cite{Leicht2006vertex} or extraction of protein pairs for a similar function from protein-protein interaction network\cite{HP}).
Also, similarity indices are applied to link prediction, which makes a priority list for missed connections in an empirically observed network\cite{Linkpred, survey}. As a practical usage, they are also applied to recommendation systems\cite{recom1,recom2}.
There are many similarity indices; therefore, we should employ a proper index that performs well in a given network.
However, it is difficult to define which similarity index performs well in an empirical system without node metadata. Therefore, the selection of a proper similarity index in a given network is a challenging problem.

There have been some trials to verify the performance of similarity indices from model networks.
Leicht \emph{et al.} employed a stratified model network to investigate the performance of a similarity index\cite{Leicht2006vertex}.
They defined a new similarity measure called the LHN index and tested it in model networks where connection probability is determined from the difference in age of node pairs, which is allocated as a 10-integer value.
They tested their index in this model network and observed a negative correlation between age difference and similarity values.
In addition, they measured Pearson correlation between age difference and similarity values as a accuracy metric.
However, their model study is hard to generalize because the relation between node property and similarity is not linear.
Thus, other metrics may be more appropriate.
Furthermore, a wider variety of conditions should be considered to investigate performances of similarity indices in various situations.

In this study, we investigated the relationship between the performance of the similarity index and structural property of a model network.
We employed a two-state random network, which is comprised of nodes with binary type and their connection probability depending on the type of node pair.
The structure of this model network reflects node type information.
We examined how well a similarity index can find the same type of node from structural information.
In addition, we investigated a parameter related with this performance, which is link prediction performance.
We observed the relationship between discrimination performance and link prediction performance for various conditions and verified this relationship through empirical networks.

\section{Methodology}

\subsection{Similarity index}
We considered ten similarity indices. We used seven local indices, which uses structural information of common neighbors: common-neighbor (CN) index\cite{CN}, Adamic-Adar (AA) index\cite{AA}, Jaccard index\cite{Linkpred}, resource allocation (RA) index\cite{RA}, preferential attachment (PA) index\cite{Linkpred}, Sorensen index\cite{Sorensen}, and hub-promoted (HP) index\cite{HP}. The other three indices were global indices, which consider the whole network structure to calculate the similarity of a node pair: Simrank\cite{Simrank}, Katz index\cite{Katz}, and LHN index\cite{Leicht2006vertex}.
All of the employed indices and their mathematical descriptions are listed in Table \ref{index_list}.
Global indices usually have tunable parameters: C in Simrank, $\beta$ in the Katz index, and $\alpha$ in the LHN index.
These parameters give weight for longer paths, which can be arbitrarily fixed values.
In this study, we set $C=0.8$, $\beta=0.01$, and $\alpha=0.97$.

\begin{table}
\begin{tabular}{c|c}
\hline
Name of similarity index & Mathematical definition  \\
\hline
Common-neighbor (CN) & $ |\Gamma (x) \cap \Gamma(y)| $ \\
Adamic-Adar (AA) & $ \sum_{z \in \Gamma (x) \cap \Gamma(y)}  \frac{1}{log (|\Gamma (z)|)}  $ \\
Jaccard & $ \frac{|\Gamma (x) \cap \Gamma(y)|}{|\Gamma (x) \cup \Gamma(y)|} $  \\
Resource Allocation (RA) & $ \sum_{z \in \Gamma (x) \cap \Gamma(y)} \frac{1}{|\Gamma (z)|} $ \\
Preferential Attachment (PA) & $ |\Gamma (x)| |\Gamma (y)| $  \\
Sorensen & $ \frac{|\Gamma (x) \cap \Gamma (y)|}{|\Gamma (x)|+|\Gamma (y)|} $ \\
Hub-Promoted (HP) & $ \frac{|\Gamma (x) \cap \Gamma (y)|}{max(|\Gamma (x)|,|\Gamma (y)|)} $ \\
\hline
Simrank & $ s(x,y) = \frac{C}{|\Gamma (x)| |\Gamma (y)|} \sum_{x'\in \Gamma (x), y'\in \Gamma (y)} s(x',y') $ \\
Katz & $ (I-\beta \mathbf{A})^{-1} - \mathbf{I} $ \\
LHN & $ 2m\lambda_1 \mathbf{D^{-1}}(\mathbf{I}-\frac{\alpha}{\lambda_1}\mathbf{A})^{-1}\mathbf{D^{-1}} $ \\
\hline
\end{tabular}
\caption{The mathematical definition of similarity indices. Here are the definitions of math symbols:
$\Gamma (x)$ is the set of neighbors for the node $x$, $\mathbf{A}$ is the adjacency matrix of the given network, $\lambda_1$ is the largest eigenvalue of the adjacency matrix, $m$ is the mean degree, and $\mathbf{D}$ is the degree matrix with $D_{ij} = k_{i} \delta_{ij}$.
Global indices (Simrank, Katz, and LHN) have parameters determining depth level of structural property:
$C$ in the Simrank, $\beta$ in the Katz, and $\alpha$ in the LHN.
In this study, we set $C=0.8$, $\beta=0.01$, and $\alpha=0.97$.}
\label{index_list}
\end{table}

\subsection{Two-state random network model}
\begin{figure}
\includegraphics[width=9cm]{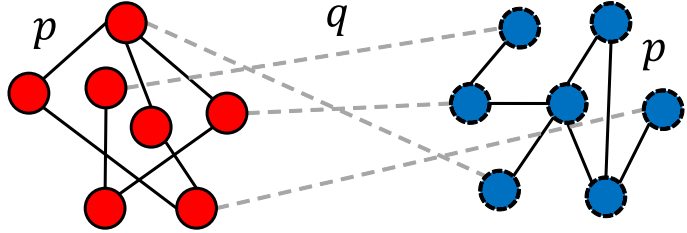}
\caption{Example of a two-state random network. Node color represents node type, which is initially given. Black lines represent connections between node pairs with the same type (intra-connection) and gray dashed lines represent those with different type (inter-connection). Connection probability is given as $p$ for intra-connection and $q$ for inter-connection.}
\label{twostate}
\end{figure}
We considered a two-state random network model, which is a simplified version of the stratified model network\cite{Leicht2006vertex} (Fig. \ref{twostate}).
In this model, two types of node exist with $N_1, N_2$ nodes between total $N$ nodes with $N=N_1+N_2$, where node type is initially given.

We consider an intra-connection to be a connection between a node pair with the same type and an inter-connection to be those with the different type.
We set connection probability for intra-connection as $p$ and probability for inter-connection as $q$.
We can obtain various structures by controlling $p$ and $q$.
When $p>q$, the model network shows community structure, and if $p<q$, the network is close to the bipartite network.
We referred to the $p>q$ case as an intra-dominant structure and the $p<q$ case as an inter-dominant structure.

We observed the relationship between performance and three structural parameters: the number of nodes, the number of links, and intra-inter ratio.
We define intra-inter ratio as $p/q$, the ratio of intra-connection probability to inter-connection probability.
When intra-inter ratio $p/q$ becomes larger or smaller than 1, node types are more strongly reflected into the network structure.
If $p/q=1$ $(p=q)$, network structure does not depend on the node types.

We set the number of nodes for each type as $N_1=N_2=N/2$ in all conditions.
We set $p/q=3$ and $6$ (intra-dominant) and $p/q=1/3$ and $1/6$ (inter-dominant) with $L=1000, 1400, 1800, 2200,$ and $2600$.
To investigate the size effect, we set network size to $N=200, 400, 600, 800,$ and $1000$ under fixed mean degrees $\langle k \rangle=10$ and $20$ and fixed intra-inter ratio at $p/q=4$ and $1/4$.
1,000 network ensembles were considered for each condition and average accuracy was obtained.

\subsection{Performance calculation}

\begin{table}
\begin{tabular}{| c | c | c |}
\hline
& Positive & Negative \\
\hline
Classified to Positive & True Positive (TP) & False Positive (FP)\\
\hline
Classified to Negative & False Negative (FN) & True Negative (TN) \\
\hline
\end{tabular}
\caption{The confusion matrix. Data have a binary type as positive or negative in a binary classification problem. A positive class corresponds to intra-connection and a negative class to inter-connection in a discrimination process. A positive class corresponds to missing link and a negative class to a nonexistent link in a link prediction process.}
\label{confusion}
\end{table}

\begin{figure}
\includegraphics[width=11cm]{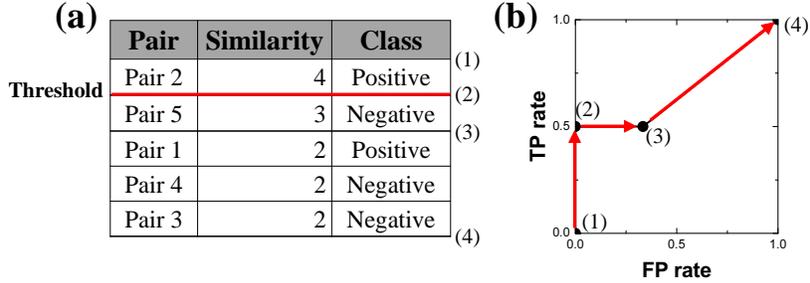}
\caption{An example of a receiver operating characteristic (ROC) curve. (a) Sample list for ROC curve example. Class refers to their originally allocated class (positive or negative). We sorted the list in descending order of similarity. None of the pairs are classified into positive class when we set the threshold in (1), so TP rate and FP rate are both 0. Thus, the starting point of the ROC curve is (0,0). We can obtain points from various thresholds: (1/2,0) from (2), (1/2,1/3) from (3), (1,1) from (4). (b) The ROC curve from obtained points.}
\label{AUCex}
\end{figure}

We employed the widely used area under the ROC curve (AUC) value as the accuracy metric in the binary classification problem\cite{HSM, ROC}. 
The receiver operating characteristic (ROC) curve is a curve in 2-dimensional space, where the x-axis corresponds to the true positive (TP) rate and the y-axis to the false positive (FP) rate:
\begin{equation}
TP rate = \frac{TP}{TP+FN},   FP rate = \frac{FP}{FP+TN},
\end{equation}
where $TP$, $TN$, $FP$, and $FN$ are described in Table. \ref{confusion}.
Pairs above a threshold of similarity index $L$ are classified in the positive class. We can obtain points from various $L$ values (Fig. \ref{AUCex} (a)). Then, we can obtain the ROC curve by connecting consecutive points (Fig. \ref{AUCex} (b)). The AUC value is the area under the ROC curve.

There are two criteria for comparing performance: a random classifier and a perfect classifier.
The AUC has a maximum value of 1. The AUC value for a random classifier is 0.5.
A poor classifier has an AUC value of less than 0.5, and a good classifier has an AUC value close to 1.
Usually, the AUC value lies between 0.5 and 1.

We calculated AUC value for two independent processes: discrimination process and link prediction process.
The discrimination process is the process that determines whether node types of a given node pair are of the same type or of different type.
In this problem, a positive class corresponds to a node pair with the same type and a negative class to one with different type.
Then, we can treat this problem as a binary classification problem.
We will call this classification performance discrimination performance.
We can also quantify the performance of link prediction processes using the AUC value.
In this process, the positive class corresponds to a missing link and the negative class to nonexistent links.

\subsection{Discrimination process and link prediction process}
\begin{figure}
\includegraphics[width=9cm]{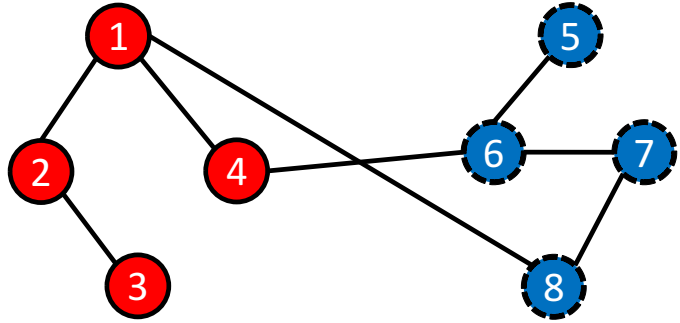}
\caption{The example for difference between discrimination and link prediction. In the discrimination, nodes 3 and 4 are considered as a positive class, but they will be treated as a nonexistent links (negative class) in link prediction. In addition, nodes 4 and 6 are considered as a negative class in discrimination, but they can be positive class in the link prediction if the connection is selected as a missing link.}
\label{disc-linkpred}
\end{figure}
We consider the discrimination process to be a classification of the node pairs into node pairs with the same type and node pairs with different type in a given network using a similarity index.
The similarity indices listed in Table \ref{index_list} were applied as discriminators, assuming node pairs with the same type have higher similarity than those with the different type.
Under the assumption, we can set the node pairs with the same type as the positive class and those with different type as the negative class (Table \ref{confusion}).
Then, we can treat this problem as a binary classification problem.
We listed all of the node pairs in a given network and sorted the list in descending order of similarity (Fig. \ref{AUCex} (a)).

Link prediction estimates probable missing links between nonadjacent pairs in the remained network from the priority list\cite{survey,HSM}.
The goal of link prediction is to find missing links from nonadjacent pairs in the remaining network using topological information of the remaining network.
Usually, there are no metadata for missed connections; therefore, we created missing links in the original network by randomly selecting links and deleting them to estimate link prediction performance\cite{HSM}.
To test the methodology, a network was divided into the remaining network ($E_t$, training set) and missing links ($E_p$, probe set) with $E_t \cup E_p = E$ and $E_t \cap E_p = \phi$ for missing link creation.
After division, nonadjacent pairs in the remaining network can be classified into two types: missing link (they are connected in the original network) and nonexistent link (they are not connected in the original network).
Then, we can treat the link prediction problem in the same way as the binary classification problem, assuming that missing links have higher similarity than nonexistent links\cite{survey,HSM}.
Similar to the discrimination process, we listed all nonadjacent pairs in the remaining network and sorted the list in descending order of similarity.
For the measurement of link prediction performances, we considered 200 independent missing link creation steps.
5\% of links were removed and considered as missing links.

The two processes are quite similar because they both employ a similarity index and make a priority list for discrimination, but they are different in detail (Fig. \ref{disc-linkpred}).
In the discrimination process, we calculated similarity from the original network structure and consider all of the node pairs.
However, the link prediction process deletes some links to create missing links, and similarity is calculated on the remaining network.
In addition, link prediction only considers nonadjacent pairs in the remaining network.
Therefore, the same node pair can be classified into different classes.
Positive pairs in the discrimination process can be negative pairs in the link prediction process and vice versa (Fig. \ref{disc-linkpred}).

We can employ the same similarity indices for both the discrimination process and the link prediction process.
To reveal the relationship between them, we obtained the performance of them for each index.
We observed a scatter plot for each pair’s performance of indices for one sample model network.

\section{Results and discussion}

\subsection{Correlation between discrimination performance and structural parameters}
\begin{figure}
\includegraphics[width=14.5cm]{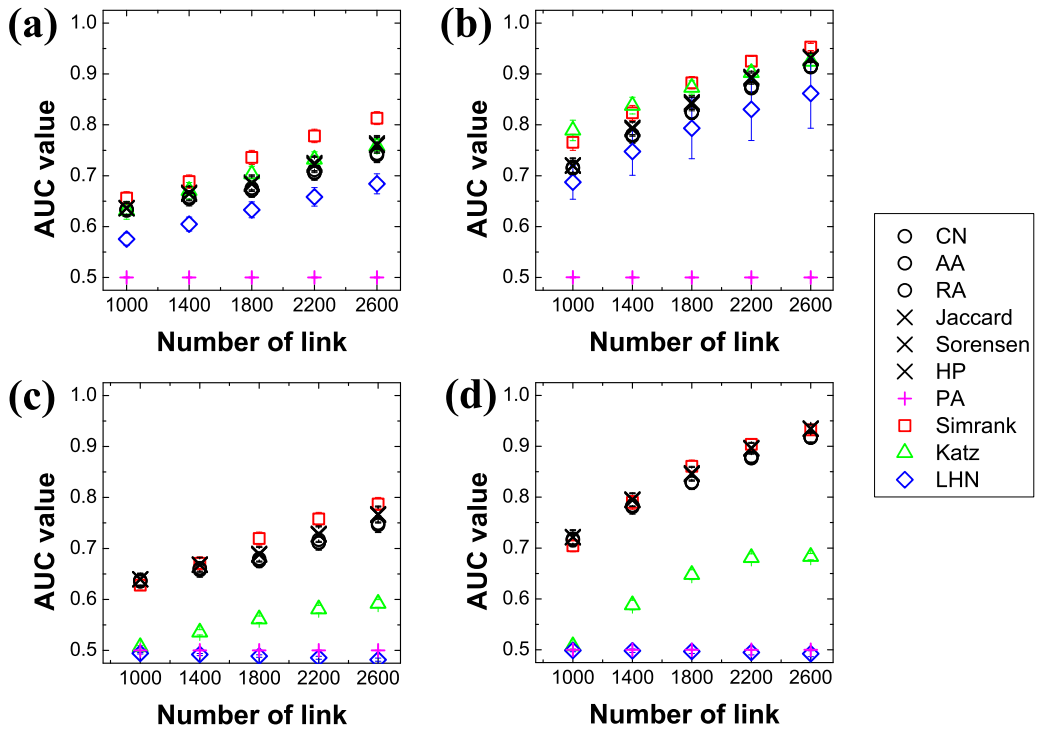}
\caption{Discrimination performance for (a) $p/q=3$, (b) $p/q=6$, (c) $p/q=1/3$, and (d) $p/q=1/6$ with $N=200$. Generally, performance increased with increased numbers of links and increased intra-inter ratio $p/q$, except for PA in all cases and some global indices (Katz and LHN) in inter-dominant cases. Some local indices that showed similar performance are represented by the same symbols.}
\label{intra-dominant}
\end{figure}

\begin{figure}
\includegraphics[width=14.5cm]{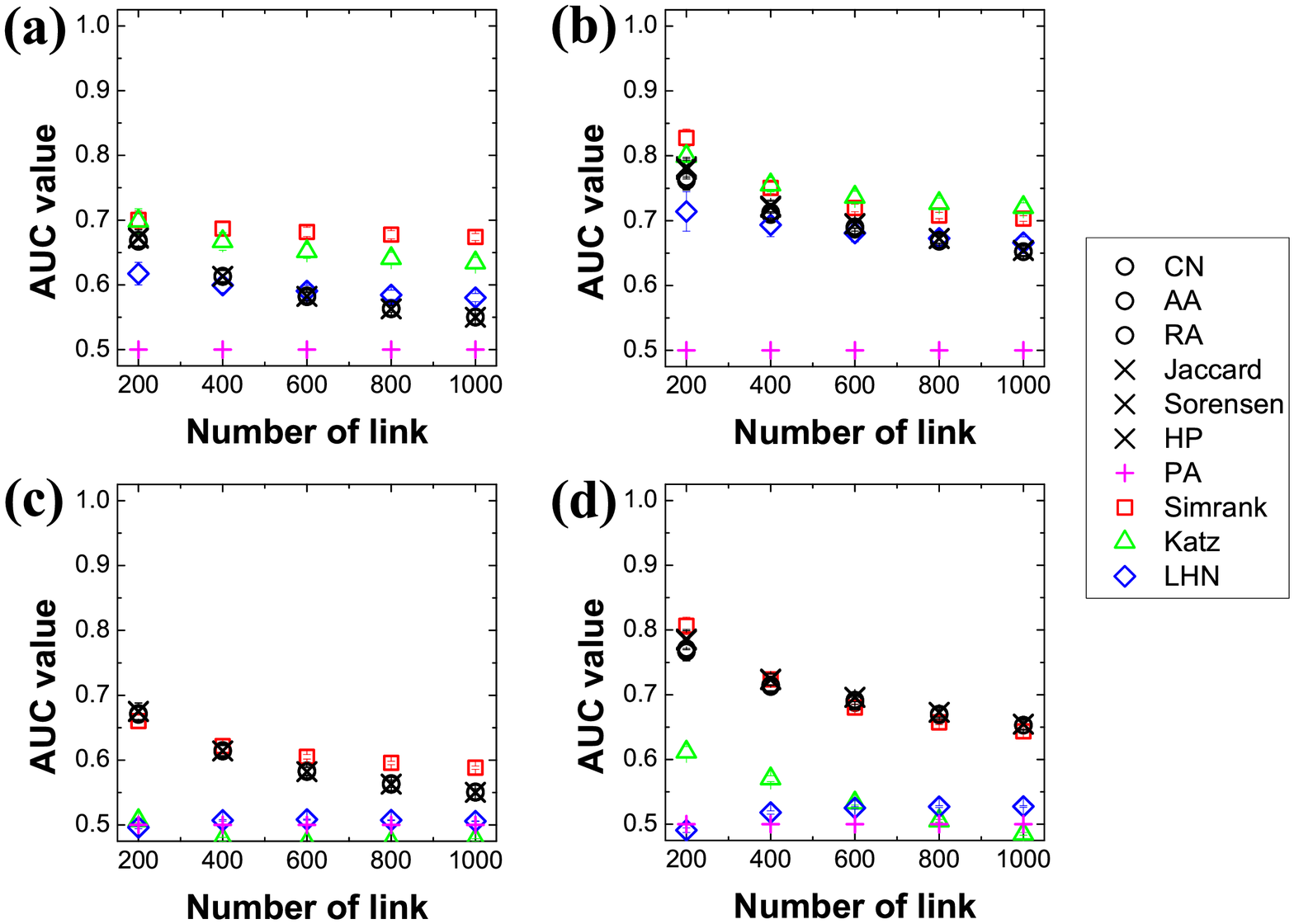}
\caption{Discrimination performance for (a) $\langle k \rangle = 10$, $p/q=4$, (b) $\langle k \rangle = 20$, $p/q=4$, (c) $\langle k \rangle = 10$, $p/q=1/4$, and (d) $\langle k \rangle = 20$, $p/q=1/4$. In the small sized network, local similarity indices performed well, but global indices showed superior performance with larger sized networks in intra-dominant condition. Some local indices that showed similar performance are represented by the same symbols.}
\label{size}
\end{figure}

We observed a positive correlation between link density and discrimination performance in both intra-dominant and inter-dominant structures (Fig. \ref{intra-dominant}).
This result is related to the sparsity problem in collaborative filtering, which comes from a small amount of available structural information of the small size data\cite{recom2}.
Intra-inter ratio is also important factor for discrimination performances.
Discrimination performances are increased when ratio is far from 1, with the same number of links.
Similarity indices perform better when the structural gap between intra-connection and inter-connection becomes larger, regardless of whether intra-dominant or inter-dominant.
However, some global indices (Katz and LHN) show poor performance in inter-dominant structures.
These indices are based on the number of paths between node pairs.
Connected pairs has higher similarity than unconnected pairs because of decay factor of these indices.
Thus, there are more node pairs with the different type with higher similarity in 'inter-dominant condition', leading to poor performances.

All indices show negative correlation with network size, but global indices are less dependent on the network size in intra-dominant structures.
Previous studies reported that global indices perform better than local indices because of their higher computational complexity\cite{Linkpred,RA}.
However, in intra-dominant condition, local indices show similar performance and even exceed the performance of some global indices in dense, small-size networks.
Nevertheless, global indices can be a better choice in large, sparse networks.
The performance of global indices exceeds that of all the local indices in a large network size (Fig. \ref{size}), and their difference becomes larger with lower mean degree.
However, although global indices are better than local indices with a large network size, absolute performances of them are lower than with a small network size.
Lack of structural information may affect the discrimination performances.
Some different behaviors appear in inter-dominant structures.
Local simiarity indices perform better even in large, dense network.
In addition, negative correlation also holds in all of indices except the LHN index.
Connected pairs have higher LHN index values than unconnected pairs, therefore connected pair with different type shows more larger in the intra-dominant conditions.
However, similarity of unconnected pair shows no much different between the pair with the same type and those with different type.
The proportion of connected pairs becomes smaller for larger network size, which causes positive correlation.
However, this positive correlation does not mean LHN can be a good choice in inter-dominant large networks.
Their performance is close to 0.5, which is worse than all of the local indices.

Can similarity indices catch `similar' nodes compared with random classifier?
In the intra-dominant condition, the answer is yes, except for with PA, and in the inter-dominant condition, the answer is partly yes.
Local indices and Simrank show better performance in the inter-dominant condition.
From this result, we can deduce that some similarity indices can generally catch similar nodes in both inter-dominant and intra-dominant structures.
However, some global indices show a lower performance in the inter-dominant condition.
Characteristics of similarity indices should be considered when we apply them.

Some indices showed similar performances; CN, AA, and RA were similar in performance, and so to were Jaccard, Sorensen, and HP.
Their performances were so close that each trio appear as a single point in Fig. \ref{intra-dominant} and Fig. \ref{size}.
The mathematical definitions of these clustered indices have some common features:
CN, AA, and RA are defined from the properties of common neighbors, and Jaccard, Sorensen, and HP are defined from the number of common neighbors and appropriate normalization term.
Their mathematical analogousness may be reflected in their discrimination performances.

\subsection{Correlation between discrimination performance and link prediction performance}
\begin{figure}
\includegraphics[width=13cm]{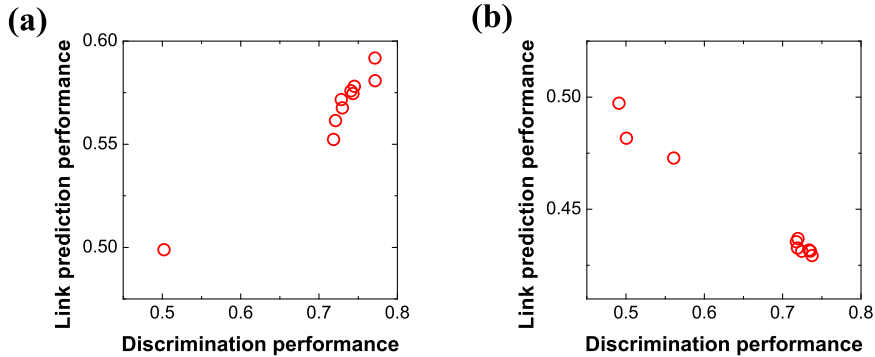}
\caption{Correlation between discrimination performance and link prediction performance under (a) intra-dominant and (b) inter-dominant structure. Figures are obtained from one sample of a two-state random network with $N=200$ and (a) $p=0.12$ and $q=0.03$ and (b) $p=0.03$ and $q=0.12$.}
\label{correlation-model}
\end{figure}

We discussed the relationship between discrimination performance and network parameters in the previous chapter, so how do we select proper similarity indices for given networks when we cannot access node metadata?
From the relationship between discrimination performance and link prediction performance, we have revealed some clues about this problem.

We calculated discrimination performance and link prediction performance for one sample model network for all indices.
Each index can be used in both the discrimination process and link prediction process.
Then, we can obtain a performance pair for each index.
We obtained a scatter plot from them for a single ensemble of model networks with intra-dominant and inter-dominant structures, where a point in the scatter plot represents the performance pair of a similarity index.

Fig. \ref{correlation-model} shows that discrimination performance and link prediction performance are correlated.
However, their correlation relationship depends on their structural characteristics:
positive correlation when intra-dominant, and negative correlation when inter-dominant.
In the intra-dominant structure, intra-connections are mainly selected in the missing link creation process.
Therefore, good discrimination performance ensures higher similarity for missing links.
However, when the inter-dominant structure, inter-connections are usually selected in the missing link creation process.
Thus, good discrimination performance leads to lower similarity values for missing links because intra-connection has higher similarity, even in the inter-dominant structure.

\begin{table}
\begin{tabular}{ccccccc}
\hline
Network & N & $N_1$ & $N_2$ & L & $L_{Intra}$ & $L_{Inter}$ \\
\hline
Karate & 34 & 17 & 17 & 78 & 67 & 11 \\
Adjnoun & 112 & 58 & 54 & 425 & 119 & 306 \\
\hline
\end{tabular}
\caption{Network properties of employed empirical networks. $L_{intra}$ is the number of intra-connection and $L_{inter}$ is the number of inter-connection. Karate shows intra-dominant structure and Adjnoun shows inter-dominant structure.}
\label{empirical_property}
\end{table}
\begin{figure}
\includegraphics[width=13cm]{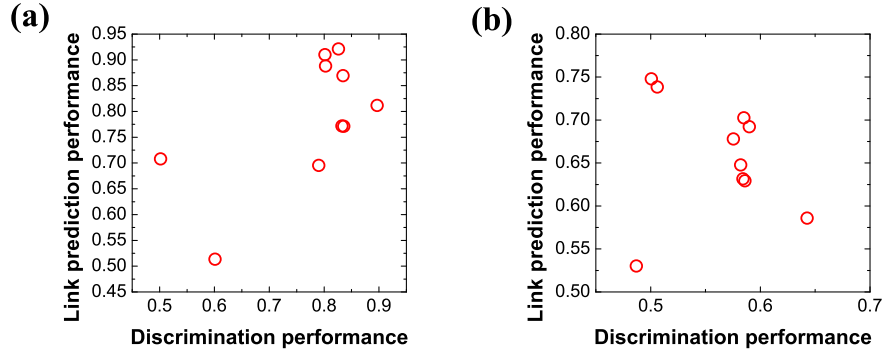}
\caption{Empirical investigation of the correlation between discrimination performance and link prediction performance for (a) Karate and (b) Adjnoun.}
\label{correlation-empirical}
\end{figure}

We verified the relationship between discrimination performance and link prediction performance from empirical networks with binary types.
Two empirical networks are employed: the social network of a karate club in a university, where node type represents their social group\cite{karate} (called Karate) and a word adjacency network from the novel \emph{David Copperfield} by Charles Dickens, where nodes are words and node types are noun and adjacent and a node pair is connected when both nodes are adjacent in the corpus\cite{adjnoun} (called Adjnoun).
Their network properties are listed in Table \ref{empirical_property}.
A positive correlation is observed in Karate, whereas Adjnoun shows a negative correlation, which is consistent with the results of the models.

Discrimination performance cannot be calculated when metadata about node type are unavailable.
However, link prediction performance requires only structural information of the network and can therefore be obtained without metadata.
We expect to approximate discrimination performance without metadata from the linear correlation with link prediction performance.

\subsection{Discussion}
We investigated the discrimination performance of model networks, where node type and structural property reflects equivalence of node types.
In the model test, results for the relationship between network parameter and performance are also observed in the link prediction process\cite{Ahn}.
However, similarity indices perform well even in inter-dominant structures except for some global indices.
This observation shows that these similarity indices can detect similar nodes regardless of intra-dominant or inter-dominant structure.
The characteristics of the local indices and global indices are different.
Performance of the local indices are clustered and shows consistent behaviors.
Performance of each global index, however, shows different characteristics.
Therefore, a deeper understanding of the characteristics of a given network will be required when we use global indices.
Global indices may not a good choice when given network is close to inter-dominant structure.

In addition, we observed a correlation between discrimination performance and link prediction performance.
Link prediction performance is obtained from structural information with no meta-information. Therefore, link prediction performance is expected to be an approximation of the discrimination performance when metadata of the node is inaccessible.
However, this relationship depends on whether network structure is intra-dominant or inter-dominant. Therefore, development by prediction of correlation patterns may improve our findings as a precursor of discrimination performances.

We studied a network model with binary node type as the simplest model.
However, node types are usually various and can be continuous variables.
A wider variety of models and accuracy metrics should be investigated in a future study.

\section{CONCLUSIONS}
In this study, we investigated the relationship between discrimination performance and the structural property of a two-state random network model.
As a result, we found that the number of links and the intra-inter ratio are positively correlated with the discrimination performances of all indices.
However, their characteristics are different in detail.
Local indices show better performance in dense, small-size networks and perform well whether intra-dominant or inter-dominant.
In contrast, global indices perform better in sparse, large-size networks and some indices do not work well in inter-dominant structures.
In addition, clustered behavior of some local similarity indices is observed, which reflects the properties of their mathematical definitions.

We investigated the correlation between discrimination performance and link prediction performance.
They are positively correlated when networks are of intra-dominant structure and negatively correlated for inter-dominant structures.
This correlation is also observed in an empirical network with binary types.
Link prediction requires no metadata for node types; therefore, link prediction performance is expected to be an approximation of discrimination performance when metadata are inaccessible.

\begin{acknowledgments}

\end{acknowledgments}

\end{document}